
\magnification=1200
\input macro1.tex                

\font\bigbf=cmbx10 scaled\magstep1

\input macro2.tex                

\def\fig#1#2{\sspacing\hangindent=.75truein
\noindent \hbox to .75truein{Fig.\ #1.\hfil}#2
\hspacing\vskip 10pt}


\def\calM#1{{\cal M}_{#1}}

\hspacing
%
%
\Bergenrapport{1993-05}

\titletwolinesmod{DECORATED-BOX-DIAGRAM-CONTRIBUTIONS}
{TO BHABHA SCATTERING. (II)}

\author{G\"oran F\"aldt\footnote{${}^*$}{faldt@tsl.uu.se,
osland@vsfys1.fi.uib.no}}
\address{Gustaf Werners Institut, Box 535}
{S-751\ 21 Uppsala, Sweden}
\vskip 7 pt
\centerline{and}

\author{Per Osland${}^*$}
\address{Department of Physics, University of Bergen, All\'egt.~55}
{N-5007 Bergen, Norway}

\abstract{
We evaluate, in the high-energy limit, $s\gg|t|\gg m^2\gg\lambda^2$,
the sum of amplitudes corresponding to a class of Feynman diagrams
describing two-loop virtual photonic corrections to Bhabha scattering.
The diagrams considered are box and crossed box diagrams
with a vacuum polarization insertion in one of the photon lines.
}

\vfil
\eject

\newsec{Introduction}
In earlier papers we studied virtual,
photonic corrections to Bhabha scattering
\ref{\refBFOone}{K.~S.\ Bj\o rkevoll, G.\ F\"aldt and P.\ Osland,
Nucl.\ Phys.\ {B386} (1992) 280},
\ref{\refBFOtwo}{K.~S.\ Bj\o rkevoll, G.\ F\"aldt and P.\ Osland,
Nucl.\ Phys.\ {B386} (1992) 303},
\ref{\refBFOthree}{K.~S.\ Bj\o rkevoll, G.\ F\"aldt and P.\ Osland,
Nucl.\ Phys.\ {B391} (1993) 591},
\ref{\refFOone}{G.\ F\"aldt and P.\ Osland,
Bergen Scientific/Technical Report No.\ 1993-04,
submitted to Nucl.\ Phys.\ B}.
The kinematical region considered, is that of small momentum transfers,
which is important for luminosity measurements at LEP.
The diagrams considered in the first three papers are
the six ladder-like diagrams.
In the fourth paper, we consider the box and crossed box diagrams
with an extra photon line decorating one of the fermion lines.
In the present paper, we complete our study of decorated box diagrams
by reporting the results for the box and crossed box diagrams
with a vacuum polarization insertion in one of the photon lines.
There are four diagrams in this set,
the sum of which is gauge invariant.
A representative diagram, denoted diagram (a), is shown in fig.~1.
A second diagram, denoted ${\rm(a')}$, is the crossed (a) diagram,
obtained by crossing the photon line and the one-loop-corrected
photon line.
Then there are two diagrams, (b) and ${\rm(b')}$,
obtained from diagrams (a) and ${\rm(a')}$ by applying the photon self-energy
correction to the other photon line.

The Feynman amplitude for diagram (a) of fig.~1 is given by
the expression
$$\eqalignno{
\calM{(a)}=&\int{\d^4k\over(2\pi)^4} \;
\bar u(p_1')(-ie\gamma_\lambda)
{i(\rlap/ p_1-\rlap/ k +m) \over (p_1-k)^2-m^2+i\epsilon}
(-ie\gamma_\mu) u(p_1) \cr
\times&\bar v(p_2)(-ie\gamma^\nu)
{i(-\rlap/ p_2-\rlap/ k +m) \over (p_2+k)^2-m^2+i\epsilon}
(-ie\gamma^\lambda) v(p_2') \cr
\times&{-i\over k^2-\lambda^2+i\epsilon}\;
\bigl[ie^2\Pi^{\mu\nu}(k)\bigr]\;
{-i\over k^2-\lambda^2+i\epsilon}\;
{-i\over (p_1-p_1'-k)^2-\lambda^2+i\epsilon}, \eqalref{\eqMadef}}$$
where
$$\Pi^{\mu\nu}(k^2)
=(k^\mu k^\nu-g^{\mu\nu}k^2)\Pi_c(k^2), \eqnref{\eqvacpol}$$
is the renormalized vacuum polarization tensor.

Neglecting terms linear in $\gamma_5$ in each current,
we write the Feynman amplitude for a single diagram as {\refBFOone}
$${\cal M}={i\alpha^3\over4\pi}\, F_{00}(s,t)
[\bar u(p_1')\gamma^\mu u(p_1)]\;
[\bar v(p_2)\gamma_\mu v(p_2')] .
\eqnref{\eqFtwoloop}$$

We find that, asymptotically, for
$\lambda^2\ll m^2\ll|t|\ll s$,
and when we sum the contributions from the four vacuum-polarization
decorated box and crossed box diagrams,
$$F_{00}^{(\rm sum)}(s,t)={16\pi i\over 3t}
\biggl[{1\over2}\log^2{m^2\over|t|}
+\log{m^2\over|t|}\,\log{\lambda^2\over|t|}
+{5\over3}\log{\lambda^2\over|t|} +{5\over3}\log{m^2\over|t|}
-{\pi^2\over6} +{28\over9} \biggr].
\eqno(\nr)$$
In this result we have discarded terms that vanish as $1/s$
for large values of $s$.
The leading term of each individual Feynman amplitude
is proportional to $\log s$.
However, the gauge-invariant sum of the four amplitudes does not
contain such a
logarithmic dependence on $s$.
\newsec{The vacuum polarization tensor}
The structure of the vacuum polarization tensor, eq.~{\eqvacpol},
is determined by gauge invariance.
The scalar function $\Pi_c(k^2)$, with the renormalization condition
$\Pi_c(k^2=0)=0$, is given by
$$e^2\Pi_c(k^2)=-{2\alpha\over\pi}\int_0^1\d z\, z(1-z)\,
\log\biggl[1-{k^2z(1-z)\over m^2-\ieps}\biggr]. \eqno(\nr)$$
Integrating by parts, we can rewrite this familiar expression
in a more convenient form, as
$$e^2\Pi_c(k^2)={\alpha\over\pi}\int_0^1\d z\, z^2
(1-{\textstyle{8\over3}}z +{\textstyle{4\over3}}z^2)
{k^2\over k^2z(1-z)-m^2 +\ieps}. \eqno(\nr)$$
In this expression, the factor of $k^2$ in the numerator, together
with the factor of $k^2$ in eq.~{\eqvacpol}, cancel the two photon
propagators in this part of the diagram of fig.~1.
The remaining denominator has a structure similar to a single photon
propagator.
\newsec{The uncrossed diagram}
We split the amplitude {\eqMadef} into two terms,
following the decomposition of eq.~{\eqvacpol},
$${\cal M}_{\rm(a)}
={\cal M}_{\rm(a1)}+{\cal M}_{\rm(a2)}. \eqno(\nr)$$
They will be discussed separately in the two subsections below.
\newsubsec{THE FIRST PART}
The first part of the amplitude for diagram (a), eq.~{\eqMadef},
is given by the formula
$$\eqalignno{
\calM{(a1)}=&\int{\d^4k\over(2\pi)^4} \;
\bar u(p_1')(-ie\gamma_\lambda)
{i(\rlap/ p_1-\rlap/ k +m) \over (p_1-k)^2-m^2+i\epsilon}
(-ie\rlap/k) u(p_1) \cr
\times&\bar v(p_2)(-ie\rlap/k)
{i(-\rlap/ p_2-\rlap/ k +m) \over (p_2+k)^2-m^2+i\epsilon}
(-ie\gamma^\lambda) v(p_2') \cr
\times&{-i\over k^2-\lambda^2+i\epsilon}\;
\bigl[ie^2\Pi_c(k^2)\bigr]\;
{-i\over k^2-\lambda^2+i\epsilon}\;
{-i\over (p_1-p_1'-k)^2-\lambda^2+i\epsilon}. \eqalref{\eqpartone} }$$

Due to gauge invariance,
this term will not contribute when we add the corresponding one
from the crossed diagram.
This is easily understood as follows.
Employing the Dirac equation for free spinors, we can rewrite the
factor associated with the upper fermion line as
$$\eqalignno{
&\bar u(p_1')\gamma_\lambda \, {1\over \rlap/ p_1-\rlap/ k -m}\,
\rlap/ k\, u(p_1) \cr
&=\bar u(p_1')\gamma_\lambda \, {1\over \rlap/ p_1-\rlap/ k -m}\,
\bigl[(\rlap/ p_1-m) -(\rlap/ p_1-\rlap/ k -m)\bigr]u(p_1) \cr
&=-\bar u(p_1')\gamma_\lambda \, u(p_1). \eqalref{\equpper}}$$
Similarly, for the lower fermion line, we have
$$\bar v(p_2)\, \rlap/ k\, {1\over -\rlap/ p_2-\rlap/ k -m}\,
\gamma^\lambda \, v(p_2')
=-\bar v(p_2)\, \gamma^\lambda \, v(p_2'). \eqnref{\eqlower}$$
For the crossed diagram, $({\rm a}')$, the expression associated with
the upper fermion line is
identical to the one for the uncrossed diagram, eq.~{\equpper}.
For the lower fermion line, however, the expression {\eqlower}
is replaced by
$$\bar v(p_2)\, \gamma^\lambda \, {1\over -\rlap/ p_2'+\rlap/ k -m}\,
\rlap/ k\, v(p_2')
=\bar v(p_2)\, \gamma^\lambda \, v(p_2'). \eqno(\nr)$$
Since all other factors in the integrand {\eqpartone} remain unchanged
under crossing,
we conclude that $\calM{(a1)}$ is cancelled identically by the corresponding
contribution from the crossed diagram.
\newsubsec{THE SECOND PART}
The second term of eq.~{\eqMadef} is readily evaluated using the
techniques described in refs.~{\refBFOone}--{\refBFOthree}.
We start out by decomposing it as
$${\cal M}_{\rm(a2)}
=(-ie)^4(M^\gamma)_{\rho\sigma}(M_q)^{\rho\sigma},
\eqno(\nr)$$
where
$$(M^\gamma)_{\rho\sigma}
=[\bar u(p_1')\gamma_\lambda \gamma_\rho \gamma_\mu u(p_1)]\,
[\bar v(p_2)\gamma^\mu \gamma_\sigma \gamma^\lambda v(p_2')].
\eqnref{\eqMgamma}$$
Neglecting terms proportional to $m^2$, we have
$$\eqalignno{
(M_q)^{\rho\sigma}
&={\alpha\over\pi}\int_0^1\d z\, z^2
(1-{\textstyle{8\over3}}z +{\textstyle{4\over3}}z^2) \cr
&\null\times\int{\d^4k\over(2\pi)^4} \;
{(p_1-k)^\rho\over (p_1-k)^2-m^2+\ieps}\;
{(p_2+k)^\sigma\over (p_2+k)^2-m^2+\ieps} \cr
&\null\times {1\over z(1-z)k^2-m^2+\ieps}\;
{1\over(p_1-p_1'-k)^2-\lambda^2+\ieps}. &(\nr)}$$

The integral over four-momenta is a standard box-diagram integral,
which is given in sect.~3.1 of ref.~{\refFOone}.
Applying the formulas given there, we get
$$\eqalignno{
(M_q)^{\rho\sigma}
&={\alpha\over\pi}\int_0^1{\d z\over 1-z}\, z
(1-{\textstyle{8\over3}}z +{\textstyle{4\over3}}z^2) \cr
&\null\times{i\over16\pi^2}
\int_0^1\cdots\int_0^1 \d\alpha_1\cdots\d\alpha_4\,
\delta(1-\sum_{i=1}^4\alpha_i) \cr
&\null\times{1\over \Lambda(\alpha)^2}
\biggl[{k_1^\rho k_2^\sigma\over D(\alpha)^2}
-{g^{\rho\sigma}\over 2D(\alpha)}\biggr], \eqalref{\eqMq}}$$
where
$$\eqalignno{
\Lambda(\alpha)&=\alpha_1+\alpha_2+\alpha_3+\alpha_4, &(\nr) \cr
D(\alpha)&=\alpha_1\alpha_2s +\alpha_3\alpha_4t
-\biggl[(\alpha_1+\alpha_2)^2
+{\alpha_3\Lambda(\alpha)\over z(1-z)}\biggr]m^2
-\alpha_4\Lambda(\alpha)\lambda^2 \cr
&\equiv \alpha_1\alpha_2s +d, &(\nr) \cr
k_1&=(\alpha_2+\alpha_3+\alpha_4)p_1
+\alpha_2p_2 -\alpha_4Q, &(\nr) \cr
k_2&=\alpha_1 p_1 +(\alpha_1+\alpha_3+\alpha_4)p_2 +\alpha_4Q, &(\nr)}$$
and
$Q=p_1-p_1'$.

Contracting the spinor factor {\eqMgamma} with $(M_q)^{\rho\sigma}$
of eq.~{\eqMq}, and neglecting $\gamma_5$ parts of the currents,
we obtain an amplitude of the form {\eqFtwoloop} with
$$\eqalignno{
F_{00}(s,t)
&=4\int_0^1{\d z\over 1-z}\, z
(1-{\textstyle{8\over3}}z +{\textstyle{4\over3}}z^2) \cr
&\null\times\int_0^1\cdots\int_0^1 \d\alpha_1\cdots\d\alpha_4\,
\delta(1-\sum_{i=1}^4\alpha_i) \cr
&\null\times{1\over \Lambda(\alpha)^2}
\biggl[{N_{II}\over D(\alpha)^2} +{N_{I}\over 2D(\alpha)}\biggr] \cr
&\equiv F^{II}(s,t)+F^{I}(s,t), \eqalref{\eqFdecomp}}$$
and
$$\eqalignno{
N_{II}&=s[2(\alpha_1+\alpha_3+\alpha_4)(\alpha_2+\alpha_3+\alpha_4)
+\alpha_1\alpha_2], &(\nr) \cr
N_{I}&=-10. &(\nr) }$$
The numerators have been obtained by using the
Khriplovich identities
\ref{\refKhriplovich}{I.~B.\ Khriplovich, Sov.\ J.\ Nucl.\ Phys.\ 17
(1973) 298}\ as formulated in ref.~{\refBFOone}.

The term $F^{I}(s,t)$ of eq.~{\eqFdecomp} is easily shown to be of order
$1/s$, and hence negligible in our limit.
To prove this we calculate the Mellin transform of $F_I(s,t)$
with the technique described in ref.~{\refFOone}.
Putting $|t|=1$, and assuming a power $s^{-1}$ for $F_I(s,|t|=1)$, we get
$$\eqalignno{
\tilde I_I(\zeta)
&=\int_0^\infty \d s\, s^{-\zeta}\, F_I(s,|t|=1) \cr
&=-40\, {\Gamma(1-\zeta)\Gamma(\zeta)\over\Gamma(1)}
\int_0^1{\d z\over 1-z}\, z
(1-{\textstyle{8\over3}}z +{\textstyle{4\over3}}z^2) \cr
&\null\times\int_0^1\cdots\int_0^1 \d\alpha_1\cdots\d\alpha_4\,
\delta(1-\sum_{i=1}^4\alpha_i) \cr
&\null\times{1\over \Lambda(\alpha)^2}
(\alpha_1\alpha_2)^{-1+\zeta}\, d^{-\zeta}. &(\nr)}$$
The most singular contribution to the integral comes from the region
$\alpha_1\simeq\alpha_2\simeq0$ and $z\simeq1$.
Applying the Cheng-Wu theorem (cf.\ sect.~2 of ref.~{\refFOone}), we can
write the integral as
$$\eqalignno{
\tilde I_I(\zeta)
&=-{40\over\zeta}
\int_0^1{\d z\over1-z}\, z
(1-{\textstyle{8\over3}}z +{\textstyle{4\over3}}z^2)
[z(1-z)]^\zeta \cr
&\null\times \int_0^\infty\d\alpha_1 \alpha_1^{-1+\zeta}
\int_0^\infty\d\alpha_2 \alpha_2^{-1+\zeta}\,
{1\over(1+\alpha_1+\alpha_2)^2} \cr
&\simeq-{40\over\zeta^3}
\int_0^1\d z\, z
(1-{\textstyle{8\over3}}z +{\textstyle{4\over3}}z^2)
(1-z)^{-1+\zeta} \cr
&={40\over3\zeta^4}. &(\nr)}$$
Inverting the Mellin transform and reintroducing the scale factor $t$,
we conclude that, to leading order in $s$,
$$F^I(s,t)={20\over9s}\, \log^3{s\over|t|}. \eqno(\nr)$$
In our approximation, this term may be discarded.

For the second term of eq.~{\eqFdecomp}, we also perform a Mellin
transform.  Assuming no inverse power of $s$ in the
dominant term of $F^{II}(s,t)$, the relevant transform to be evaluated,
becomes
$$\eqalignno{
\tilde I_{II}(\zeta)
&=\int_0^\infty \d s\, s^{-\zeta-1}\, F^{II}(s,|t|=1) \cr
&=4\,{\Gamma(1-\zeta)\Gamma(1+\zeta)\over\Gamma(2)}\,
\int_0^1{\d z\over 1-z}\, z
(1-{\textstyle{8\over3}}z +{\textstyle{4\over3}}z^2) \cr
&\null\times\int_0^1\cdots\int_0^1 \d\alpha_1\cdots\d\alpha_4\,
\delta(1-\sum_{i=1}^4\alpha_i) \cr
&\null\times {N_{II}/s\over \Lambda(\alpha)^2}\,
(\alpha_1\alpha_2)^{-1+\zeta} d^{-1-\zeta}.
\eqalref{\eqMellin}}$$
Next, we apply the Cheng-Wu theorem in the $\alpha_1$, $\alpha_2$ variables,
$$\eqalignno{
\tilde I_{II}(\zeta)
&=4\, {\Gamma(1-\zeta)\Gamma(1+\zeta)\over\Gamma(2)}\,
\int_0^1{\d z\over 1-z}\, z
(1-{\textstyle{8\over3}}z +{\textstyle{4\over3}}z^2) \cr
&\null\times\int_0^\infty \d\alpha_1 \int_0^\infty \d\alpha_2
\int_0^1 \d\alpha_3\int_0^1 \d\alpha_4\,
\delta(1-\alpha_3-\alpha_4) \cr
&\null\times {N_{II}/s\over \Lambda(\alpha)^2}\,
(\alpha_1\alpha_2)^{-1+\zeta} d^{-1-\zeta}.
&(\nr)}$$

The most singular contribution when $\zeta\simeq0$ is obtained when
$\alpha_1\simeq\alpha_2\simeq0$.
By standard procedure we obtain
$$\eqalignno{
\tilde I_{II}(\zeta)
&=-{8\over\zeta^2}
\int_0^1{\d z\over 1-z}\, z
(1-{\textstyle{8\over3}}z +{\textstyle{4\over3}}z^2) \,
\int_0^1\d\alpha_3 {1\over -d}, &(\nr)}$$
where
$$\eqalignno{
-d&=\alpha_3(1-\alpha_3)
+{\alpha_3\over z(1-z)}\,m^2 +(1-\alpha_3)\lambda^2 \cr
&\simeq(\alpha_3+\lambda^2)\biggl(1+{m^2\over z(1-z)}-\alpha_3\biggr).
&(\nr)}$$

We can now perform the $\alpha_3$ integration and obtain
$$\eqalignno{
\tilde I_{II}(\zeta)
&=-{8\over\zeta^2}
\int_0^1\d z \, z^2
(1-{\textstyle{8\over3}}z +{\textstyle{4\over3}}z^2) \cr
&\null\times {1\over z(1-z)+m^2}
\bigl\{-\log\lambda^2 -\log m^2
+\log[z(1-z)+m^2] \bigr\} \cr
&=-{8\over 3\zeta^2}
\int_0^1 \d z
\biggl[1+4z(1-z)-{1\over 1+m^2-z}\biggr] \cr
&\null\times \bigl\{-\log\lambda^2 -\log m^2
+\log(1+m^2-z) +\log z\bigr\}, &(\nr)}$$
where in the last step we have neglected terms proportional to $m^2$.
The integration over $z$ is elementary except for the occurrence of the
dilogarithmic function
\ref{\refKolbig}{K.~S.\ K\"olbig, J.~A.\ Mignaco, and E.\ Remiddi,
BIT 10 (1970) 38},
\ref{\refDevoto}{A.\ Devoto and D.~W.\ Duke,
La Rivista del Nuovo Cimento 7 (1984) 1},
coming from the product of the last term in each parenthesis.
Since $\Li2(1)=\pi^2/6$, we get
$$\tilde I_{II}(\zeta)
={8\over 3\zeta^2}
\biggl[{1\over2}\log^2m^2
+\log m^2\,\log \lambda^2
+{5\over3}\log \lambda^2 +{5\over3}\log m^2
-{\pi^2\over6} +{28\over9} \biggr].
\eqno(\nr)$$

Inverting the Mellin transform {\eqMellin}, and reinserting for $t$,
we have
$$\eqalignno{
F^{II}(s,t)
&=-{8\over3t}\log{s\over|t|} \cr
&\null\times
\biggl[{1\over2}\log^2{m^2\over|t|}
+\log{m^2\over|t|}\,\log{\lambda^2\over|t|}
+{5\over3}\log{\lambda^2\over|t|} +{5\over3}\log{m^2\over|t|}
-{\pi^2\over6} +{28\over9} \biggr]. &(\nr)}$$
This is the desired result for the diagram of fig.~1.
\newsec{Summary}
In the previous section, we have calculated the amplitude
$F_{00}^{\rm (a)}(s,t)$ for the (a) diagram.
The amplitudes for the other three diagrams mentioned
in the Introduction are obtained by the substitution rules
$$\eqalignno{
F_{00}^{\rm (b)}(s,t) &=F_{00}^{\rm (a)}(s,t), \cr
F_{00}^{\rm (a')}(s,t)&=-F_{00}^{\rm (a)}(u,t), \cr
F_{00}^{\rm (b')}(s,t)&=-F_{00}^{\rm (b)}(u,t). &(\nr)}$$
The substitution $s\to u$ implies
$\log s \to \log s +i\pi$.  Therefore, the summed contribution from the four
diagrams becomes
$$F_{00}(s,t)={16i\pi\over3t}
\biggl[{1\over2}\log^2{m^2\over|t|}
+\log{m^2\over|t|}\,\log{\lambda^2\over|t|}
+{5\over3}\log{\lambda^2\over|t|} +{5\over3}\log{m^2\over|t|}
-{\pi^2\over6} +{28\over9} \biggr], \eqno(\nr)$$
demonstrating again that the $\log s$ dependent terms cancel when
we sum over a gauge-invariant set of diagrams.

This completes the study of the decorated box diagrams.
We note that the present result contains only one power
of the infrared logarithm,
whereas the previously studied classes of diagrams of
refs.~{\refBFOone}--{\refFOone} contained two powers.

In order to complete the evaluation of virtual two-loop QED corrections to
Bhabha scattering, one needs to add also the
two-loop corrections to the one-photon-exchange diagrams.

\acknowledgment{
This research has been supported by the Research Council of Norway
and by the Swedish National Research Council.}

\vfill\eject\immediate\closeout\rfile
\baselineskip=14pt\centerline{{\bf References}}\bigskip{\frenchspacing%
\catcode`\@=11\escapechar=` %
\input refs.tmp\vfill\eject}\nonfrenchspacing
\centerline{\bf Figure captions}

\vskip 15pt
\def\fig#1#2{\hangindent=.65truein \noindent \hbox to .65truein{Fig.\ #1.
\hfil}#2\vskip 2pt}

\fig1{Vacuum-polarization insertion into the box diagram.
There are two such diagrams, and also two crossed ones.}
\bye